\documentclass[fleqn]{2020SCGE}
\begin{document}

\ensubject{subject}

%%%%%%%%%%%%%%%%%%%%%%%%%%%%%%%%%%%%%%%%%%%%%%%%%%%%%%%
%%% Authors do not modify the information below
%%% ????????????????
%%% ??????????, ????????????{}, ???????????????????
%Letter to the Editor??Article%??????
\ArticleType{Article}%??Article
\SpecialTopic{SPECIAL TOPIC: }%???????
\Year{2025}
\Month{xx}
\Vol{xx}
\No{x}
\DOI{??}
\ArtNo{000000}
\ReceiveDate{January xx, 2025}
\AcceptDate{xx xx, 2025}
%\OnlineDate{January 1, 2016}
%%%%%%%%%%%%%%%%%%%%%%%%%%%%%%%%%%%%%%%%%%%%%%%%%%%%%%%

%%% title: ????
%%%   \title{title}{title for citation}
\title{Future Cosmology: New Physics and Opportunity \\from the China Space Station Telescope (CSST)}

%%% Corresponding author: ???????
%%%   \author[number]{Full name}{{email@xxx.com}}
%%% General author: ???????
%%%   \author[number]{Full name}{}
\author[1,2,3]{Yan Gong}{{gongyan@bao.ac.cn}}
\author[1,3]{Haitao Miao}{}
\author[1,3]{Xingchen Zhou}{}
\author[1,2]{Qi Xiong}{}
\author[1,2]{Yingxiao Song}{}
\author[1,2]{Yuer Jiang}{}
\author[1,2]{\\Minglin Wang}{}
\author[1,2]{Junhui Yan}{}
\author[1,2]{Beichen Wu}{}
\author[1,2]{Furen Deng}{}
\author[1,2,4,5]{Xuelei Chen}{}
\author[6]{Zuhui Fan}{}
\author[7,8]{\\Yipeng Jing}{}
\author[7,8]{Xiaohu Yang}{}
\author[1,9]{Hu Zhan}{}

%%% Author information for page head. 
%%% ??????????????, ??????????author???
\AuthorMark{Gong Y.}%\authorcr????????

%%% Authors for citation. ????????????????
%%% ??????????????, ??????????author???
\AuthorCitation{Gong Y,, Miao H., Zhou X., et al.}

%%% Address. ???
%%%   \address[number]{Address, City {\rm Postcode}, Country}
\address[1]{National Astronomical Observatories, Chinese Academy of Sciences, 20A Datun Road, Beijing 100101, China}
\address[2]{School of Astronomy and Space Sciences, University of Chinese Academy of Sciences (UCAS), 19A Yuquan Road, Beijing 100049, China}
\address[3]{Science Center for China Space Station Telescope, National Astronomical Observatories, Chinese Academy of Sciences, \\20A Datun Road, Beijing 100101, China}
\address[4]{Department of Physics, College of Sciences, Northeastern University, Shenyang 110819, China}
\address[5]{Centre for High Energy Physics, Peking University, Beijing 100871, China}
\address[6]{South-Western institute for Astronomy Research, Yunnan University, Kunming 650500, China}
\address[7]{Department of Astronomy, School of Physics and Astronomy, Shanghai Jiao Tong University, Shanghai 200240, China}
\address[8]{Tsung-Dao Lee Institute and Key Laboratory for Particle Physics, Astrophysics and Cosmology, Ministry of Education, Shanghai 201210, China}
\address[9]{Kavli Institute for Astronomy and Astrophysics, Peking University, Beijing 100871, China}

%\contributions{}%????????

%%% Abstract. ??
\abstract{The China Space Station Telescope (CSST) is the next-generation Stage~IV survey telescope. It can simultaneously perform multi-band imaging and slitless spectroscopic wide- and deep-field surveys in ten years and an ultra-deep field (UDF) survey in two years, which are suitable for cosmological studies. Here we review several CSST cosmological probes, such as weak gravitational lensing, two-dimensional (2D) and three-dimensional (3D) galaxy clustering, galaxy cluster abundance, cosmic void, Type Ia supernovae (SNe Ia), and baryonic acoustic oscillations (BAO), and explore their capabilities and prospects in discovering new physics and opportunities in cosmology. We find that CSST will measure the matter distribution from small to large scales and the expansion history of the Universe with extremely high accuracy, which can provide percent-level stringent constraints on the property of dark energy and dark matter and precisely test the theories of gravity.}

%%% Keywords. ?????


\maketitle

\begin{multicols}{2}
\section{Introduction}\label{sec:1}

We are currently in the era of precise cosmology. With significant improvement of observation methods and technology, our  understanding of the Universe has become more and more accurate. However, there are still several important questions that need to be answered, such as how was the Universe born? whether dark energy and dark matter exist? What is their nature? Is general relativity correct on cosmological scales? Does it need to be modified? And how? To definitively solve these problems, more powerful detection equipments and telescopes are needed.

The next-generation Stage IV survey telescopes are expected to provide such capabilities and means, such as $Euclid$ \cite{Euclid24}, Nancy Grace Roman Space Telescope (RST)\footnote{https://roman.gsfc.nasa.gov/},  Vera C. Rubin Observatory \cite{LSST19}, and the China space station telescope (CSST) \cite{Zhan11,Zhan21,Gong19}. These telescopes have large field of view and high spatial resolution and sensitivity, which will explore huge volumes of space covering vast redshift range and observe billions of galaxies. They can measure the property of the components of the Universe, formation and evolution of cosmic large-scale structure (LSS), and expansion history of the Universe with extremely high accuracy.

Among these telescopes, CSST is a space-borne telescope which uses the three-mirror anastigmat (TMA) optical system with a primary mirror aperture of 2 meters. It will launch around 2027, and orbit at an altitude of $\sim400$~km, in the same orbit as the China Manned Space Station. It contains five astronomical instruments, including the multi-band imaging and slitless spectroscopy Survey Camera (SCam), Multi-Channel Imager (MCI), Integral Field Spectrograph (IFS), Cool Planet Imaging Coronagraph (CPI-C), and THz Spectrometer (TS). Using these instruments, CSST will conduct precise detection in many areas of astronomy, such as cosmology, galaxy and stellar science, exoplanet, solar system, astrometry, and transient sources.

\begin{table*}[t]
\footnotesize
\begin{center}
\caption{CSST survey design and key parameters.}
\label{tab:CSST_para}
\vspace{1mm}
\begin{tabular}{cccccccc}
\toprule
\multicolumn{8}{c} {\bf CSST photometric surveys} \vspace{1mm}  \\
\hline
\multicolumn{1}{c} {Bands} & \multicolumn{1}{c}{$NUV$} & \multicolumn{1}{c}{$u$} &\multicolumn{1}{c}{$g$} & \multicolumn{1}{c}{$r$} & \multicolumn{1}{c}{$i$} & \multicolumn{1}{c}{$z^*$} &\multicolumn{1}{c}{$y^*$}  \\

\multicolumn{1}{c} {Wavelength range (nm)} & \multicolumn{1}{c}{$252-321$} & \multicolumn{1}{c}{$321-401$} &\multicolumn{1}{c}{$401-547$} & \multicolumn{1}{c}{$547-692$} & \multicolumn{1}{c}{$692-842$} & \multicolumn{1}{c}{$842-1080$} &\multicolumn{1}{c}{$927-1100$}  \\

\multicolumn{1}{c} {PSF size ($\le R_{\rm EE80}$/FWHM$^a$)} & \multicolumn{1}{c}{$0.15''/0.20''$} & \multicolumn{1}{c}{$0.15''/0.20''$} &\multicolumn{1}{c}{$0.15''/0.20''$} & \multicolumn{1}{c}{$0.15''/0.20''$} & \multicolumn{1}{c}{$0.16''/0.21''$} & \multicolumn{1}{c}{$0.18''/0.24''$} &\multicolumn{1}{c}{$0.18''/0.24''$}   \vspace{1mm} \\

\hline
\multicolumn{8}{c} {Wide field (17500 deg$^2$ in ten years)} \vspace{1mm}  \\

\multicolumn{1}{c} {Exposure time} & \multicolumn{1}{c}{$150$\,s $\times4$} & \multicolumn{1}{c}{$150$\,s $\times2$} &\multicolumn{1}{c}{$150$\,s $\times2$} & \multicolumn{1}{c}{$150$\,s $\times2$} & \multicolumn{1}{c}{$150$\,s $\times2$} & \multicolumn{1}{c}{$150$\,s $\times2$} &\multicolumn{1}{c}{$150$\,s $\times4$}  \\

\multicolumn{1}{c} {Mag. limit$^b$} & \multicolumn{1}{c}{$25.4$} & \multicolumn{1}{c}{$25.4$} &\multicolumn{1}{c}{$26.3$} & \multicolumn{1}{c}{$26.0$} & \multicolumn{1}{c}{$25.9$} & \multicolumn{1}{c}{$25.2$} &\multicolumn{1}{c}{$24.4$}  \vspace{1mm} \\

\hline
\multicolumn{8}{c} {Deep field (400 deg$^2$ in ten years)} \vspace{1mm}  \\

\multicolumn{1}{c} {Exposure time} & \multicolumn{1}{c}{$250$\,s $\times16$} & \multicolumn{1}{c}{$250$\,s $\times8$} &\multicolumn{1}{c}{$250$\,s $\times8$} & \multicolumn{1}{c}{$250$\,s $\times8$} & \multicolumn{1}{c}{$250$\,s $\times8$} & \multicolumn{1}{c}{$250$\,s $\times8$} &\multicolumn{1}{c}{$250$\,s $\times16$}  \\

\multicolumn{1}{c} {Mag. limit$^b$} & \multicolumn{1}{c}{$26.7$} & \multicolumn{1}{c}{$26.7$} &\multicolumn{1}{c}{$27.5$} & \multicolumn{1}{c}{$27.2$} & \multicolumn{1}{c}{$27.0$} & \multicolumn{1}{c}{$26.4$} &\multicolumn{1}{c}{$25.7$}  \vspace{1mm} \\

\hline
\multicolumn{8}{c} {Ultra-deep field (UDF, 9 deg$^2$ in two years)} \vspace{1mm} \\

\multicolumn{1}{c} {Exposure time} & \multicolumn{1}{c}{$250$\,s $\times120$} & \multicolumn{1}{c}{$250$\,s $\times60$} &\multicolumn{1}{c}{$250$\,s $\times60$} & \multicolumn{1}{c}{$250$\,s $\times60$} & \multicolumn{1}{c}{$250$\,s $\times60$} & \multicolumn{1}{c}{$250$\,s $\times60$} &\multicolumn{1}{c}{$250$\,s $\times120$}  \\

\multicolumn{1}{c} {Mag. limit$^b$ (multi/single expo.)} & \multicolumn{1}{c}{$28.0/25.3$} & \multicolumn{1}{c}{$28.0/25.7$} &\multicolumn{1}{c}{$28.7/26.4$} & \multicolumn{1}{c}{$28.4/26.1$} & \multicolumn{1}{c}{$28.2/25.9$} & \multicolumn{1}{c}{$27.7/25.4$} &\multicolumn{1}{c}{$27.1/24.4$}  \vspace{1mm} \\

\toprule
\multicolumn{8}{c} {\bf CSST spectroscopic surveys (slitless gratings)} \vspace{1mm}  \\
\hline
\multicolumn{1}{c} {Bands} & \multicolumn{2}{c}{$GU$} & \multicolumn{2}{c}{$GV$} &\multicolumn{3}{c}{$GI^*$}\\

\multicolumn{1}{c} {Wavelength range (nm)} & \multicolumn{2}{c}{$255-410$} & \multicolumn{2}{c}{$400-640$} &\multicolumn{3}{c}{$620-1000$}\\

\multicolumn{1}{c} {Spectral resolution $R=\lambda/{\Delta \lambda}$} & \multicolumn{2}{c}{$\ge 200$} & \multicolumn{2}{c}{$\ge 200$} &\multicolumn{3}{c}{$\ge 200$} \vspace{1mm}\\

\hline
\multicolumn{8}{c} {Wide field (17500 deg$^2$ in ten years)} \vspace{1mm}  \\

\multicolumn{1}{c} {Exposure time} & \multicolumn{2}{c}{$150$\,s $\times4$} & \multicolumn{2}{c}{$150$\,s $\times4$} &\multicolumn{3}{c}{$150$\,s $\times4$}\\

\multicolumn{1}{c} {Mag. limit$^b$} & \multicolumn{2}{c}{$23.2$} & \multicolumn{2}{c}{$23.4$} &\multicolumn{3}{c}{$23.2$} \vspace{1mm}\\

\hline
\multicolumn{8}{c} {Deep field (400 deg$^2$ in ten years)} \vspace{1mm}  \\

\multicolumn{1}{c} {Exposure time} & \multicolumn{2}{c}{$250$\,s $\times16$} & \multicolumn{2}{c}{$250$\,s $\times16$} &\multicolumn{3}{c}{$250$\,s $\times16$}\\

\multicolumn{1}{c} {Mag. limit$^b$} & \multicolumn{2}{c}{$24.4$} & \multicolumn{2}{c}{$24.5$} &\multicolumn{3}{c}{$24.3$} \vspace{1mm}\\

\hline
\multicolumn{8}{c} {Ultra-deep field (UDF, 9 deg$^2$ in two years)} \vspace{1mm} \\

\multicolumn{1}{c} {Exposure time} & \multicolumn{2}{c}{$250$\,s $\times120$} & \multicolumn{2}{c}{$250$\,s $\times120$} &\multicolumn{3}{c}{$250$\,s $\times120$}\\

\multicolumn{1}{c} {Mag. limit$^b$} & \multicolumn{2}{c}{$25.6$} & \multicolumn{2}{c}{$25.7$} &\multicolumn{3}{c}{$25.6$} \vspace{1mm}\\

\bottomrule
 \end{tabular}
\end{center}
\vspace{-2mm}
$^*$ cut off by detector quantum efficiency. \\
$^a$ Assuming Gaussian-like PSF profile, and $R_{\rm EE80}$ is the radius of 80\% energy concentration.\\
$^b$ For 5$\sigma$ point source detection in AB mag. For extend sources, e.g. galaxies, the magnitude limit can be $\sim1$ mag shallower \cite{Gong19}.\\
%%\label{tab:results}
\end{table*}

The CSST SCam will perform 17500 deg$^2$ wide-field and 400 deg$^2$ deep-field multi-band imaging and spectroscopic surveys simultaneously in ten years, and conduct a 9 deg$^2$ ultra-deep field (UDF) survey in the first two years. The field of view of SCam is $\sim1.1$ deg$^2$ with a spatial resolution $\sim0.15''$ (80\% energy concentration region), and the wavelength coverage is from $\sim$255 to 1000 nm. It has seven photometric bands, i.e. $NUV$, $u$, $g$, $r$, $i$, $z$, and $y$, with magnitude limit $i\simeq25.9$ for $5\sigma$ point source detection in the wide field survey, and three spectroscopic bands, i.e. $GU$, $GV$, and $GI$, with magnitude limit $GV\simeq23.4$ and spectral resolution $R\gtrsim200$. The details of CSST survey design and key parameters can be found in Table~\ref{tab:CSST_para}. Therefore, it has significant advantages in the study of cosmology, and is expected to yield many important discoveries by implementing different cosmological probes.

In this work, we discuss the capability of CSST in the studies of the property dark energy and dark matter, the theories of modified gravity, and other important cosmological topics, by measuring cosmic matter distribution and expansion history of the Universe with  weak gravitational lensing, galaxy clustering, galaxy clusters, cosmic voids, Type Ia supernovae (SNe Ia), baryonic acoustic oscillations (BAO), etc. The paper is organized as follows: we discuss the cosmic matter distribution measurements by CSST in Section 2, and cosmic expansion history or distance measurements in Section 3. In Section 4, we discuss and summarize the results.

\section{Cosmic matter distribution measurements}\label{sec:2}

In this section, we discuss the related CSST measurements of the matter distribution or structure of the Universe, and explore their constraints on the models of dark energy, dark matter, and modified gravity.

\subsection{Weak gravitational lensing}

Weak gravitational lensing is one of the main cosmological probes of the CSST multi-band imaging photometric survey. Since CSST  has high spatial resolution and Gaussian-like point-spread function (PSF), it is very suitable for the galaxy shape measurement in weak gravitational lensing \cite{Gong19}. Besides, the CSST has seven photometric bands from near-ultraviolet (UV) to near-infrared (IR) bands, which is expected to obtain high-precision photometric redshift estimation. All of these can effectively suppress the systematics in the weak lensing observation, and improve the measurement accuracy.

\begin{figure*}
    \centering
    \includegraphics[width=1 \linewidth]{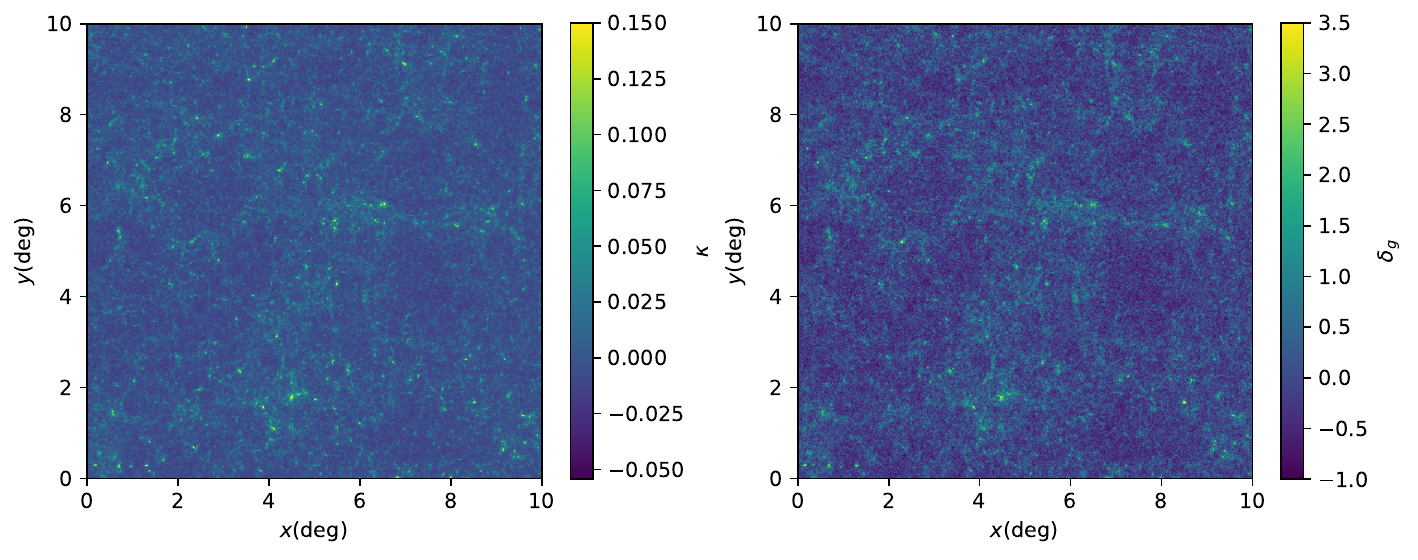}
    \caption{The CSST reconstructed convergence map (left panel) and galaxy density map (right panel) in 100 deg$^2$ from Jiutian simulations. The convergence and galaxy density maps show the corresponding signals at $0.55<z<0.89$ and $0<z<0.55$, respectively.}
    \label{fig:WL_map}
\end{figure*}

Usually, the signal of weak lensing can be measured by the two-point shear correlation functions with angle separation $\theta$, i.e. $\xi_{\pm}^{ij}(\theta)$, for the $i$th and $j$th tomographic photometric redshift (photo-$z$) bins, which is given by
\begin{equation}
\xi_{\pm}^{ij}(\theta) = \int \frac{{\rm d}\ell\, \ell}{2\pi}\,\tilde{C}_{\gamma\gamma}^{ij}(\ell)\,J_{0,4}(\ell\theta),
\label{eq:xi_pm}
\end{equation}
where $\ell$ is the multipole, and $J_{0,4}(\ell\theta)$ is the zeroth order or fourth order Bessel function of the first kind for $\xi_{+}$ and $\xi_{-}$, respectively. $\tilde{C}_{\gamma\gamma}^{ij}(\ell)$ is the measured shear power spectrum considering the intrinsic alignments and systematics, which can be written as
\begin{equation}
\tilde{C}_{\gamma\gamma}^{ij}(\ell) = (1+m_i)(1+m_j)\, {C}_{\gamma\gamma}^{ij}(\ell) + \delta_{ij}\frac{\sigma^2_{\gamma}}{\bar{n}_i} + N_{\rm add}^{ij}.
\end{equation}
Here $m_i$ is the multiplicative error in the $i$th redshift bin, which is one of the main systematics in shear calibration, $\delta_{ij}$ is the Kronecker delta function, $\sigma^2_{\gamma}=0.04$ is the shear variance per component, and $\bar{n}_i$ is the average galaxy surface number density in the $i$th tomographic bin per steradian. For the CSST photometric survey, the total surface galaxy number density is $\sim 28$ arcmin$^{-2}$ \cite{Gong19}. $N_{\rm add}^{ij}$ is additive error, which can be well controlled within $10^{-10}$ in the Stage IV weak lensing surveys \cite{Heymans06,Massey07,Massey13,Kitching12}. ${C}_{\gamma\gamma}^{ij}(\ell)$ is the shear signal power spectrum, and it is expressed as
\begin{equation}
{C}_{\gamma\gamma}^{ij}(\ell) = P_{\kappa}^{ij}(\ell) + {C}_{\rm II}^{ij}(\ell) + {C}_{\rm GI}^{ij}(\ell),
\end{equation}
where $P_{\kappa}^{ij}(\ell)$ is the convergence power spectrum, and ${C}_{\rm II}^{ij}(\ell)$ and ${C}_{\rm GI}^{ij}(\ell)$ are the intrinsic-intrinsic and gravitational-intrinsic power spectra, respectively, which accounts for the correlation of the intrinsic ellipticities of neighboring galaxies, and the correlation of a foreground galaxy and the gravitational shear of a background galaxy \cite{Joachimi15}. 

Assuming the flat-sky and Limber approximation \cite{Limber54}, $P_{\kappa}^{ij}(\ell)$ can be expressed as
\begin{equation}
P_{\kappa}^{ij}(\ell) = \int_0^{\chi_{\rm H}} {\rm d}\chi\, \frac{W^i_{\kappa}(\chi)W^j_{\kappa}(\chi)}{r^2(\chi)}\, P_{\rm m}\left( k=\frac{\ell+1/2}{r(\chi)},\chi\right),
\end{equation}
where $\chi$ is the comoving radial distance, $\chi_{\rm H}$ is the horizon distance, $r(\chi)$ is the comoving angular diameter distance, $k$ is the wavenumber, and $P_{\rm m}$ is the matter power spectrum. $W^i_{\kappa}(\chi)$ is the lensing weighting function in the $i$th tomographic bin, which is given by
\begin{equation}
W^i_{\kappa}(\chi) = \frac{3\Omega_{\rm m}H_0^2}{2c^2}\frac{r(\chi)}{a(\chi)}\int_{\chi}^{\chi_{\rm H}} {\rm d}\chi'\, n_i(\chi')\frac{r(\chi'-\chi)}{r(\chi')}.
\end{equation}
Here $\Omega_{\rm m}$ is the matter density parameter, $H_0$ is the Hubble constant, $c$ is the speed of light, $a$ is the scale factor, and $n_i$ is the normalized galaxy distribution of the $i$th tomographic bin that we have $\int n_i(\chi)\,{\rm d}\chi=1$. The intrinsic-intrinsic and gravitational-intrinsic power spectra are
\begin{equation}
C_{\rm II}^{ij}(\ell) = \int_0^{\chi_{\rm H}} {\rm d}\chi \, \frac{n_i(\chi)n_j(\chi)}{r^2(\chi)}\, P_{\rm II}\left( \frac{\ell+1/2}{r(\chi)},\chi\right),
\end{equation}
\begin{equation}
C_{\rm GI}^{ij}(\ell) = \int_0^{\chi_{\rm H}} {\rm d}\chi \, \frac{n_i(\chi)W^i_{\kappa}(\chi)+n_j(\chi)W^j_{\kappa}(\chi)}{r^2(\chi)}\, P_{\rm GI}\left( \frac{\ell+1/2}{r(\chi)},\chi\right),
\end{equation}
where $P_{\rm II}=F^2(\chi)P_{\rm m}$ and $P_{\rm GI}=F(\chi)P_{\rm m}$, and the factor $F(\chi)$ in the $i$th tomographic bin is given by
\begin{equation}
F_i(\chi) = -A_{\rm IA}C_1\rho_{\rm c} \frac{\Omega_{\rm m}}{D(\chi)}\left( \frac{1+z}{1+z_0} \right)^{\eta_{\rm IA}}\left( \frac{L_i}{L_0}\right)^{\beta_{\rm IA}}.
\end{equation}
Here $A_{\rm IA}$, $\eta_{\rm IA}$, and $\beta_{\rm IA}$ are the free parameters, and $\beta_{\rm IA}$ is usually set to be 0 to ignore the dependence of luminosity. $C_1=5\times 10^{-14}\, h^{-2}M_{\odot}^{-1}{\rm Mpc}^3$ is the normalization constant, $\rho_{\rm c}$ is the present critical density, $D(\chi)$ is the growth factor normalized to unity at $z=0$, $L_i$ is the weighted average luminosity in the $i$th tomographic bin, and $z_0$ and $L_0$ are the pivot redshift and luminosity. 

\begin{figure}[H]
\centering
\includegraphics[scale=0.55]{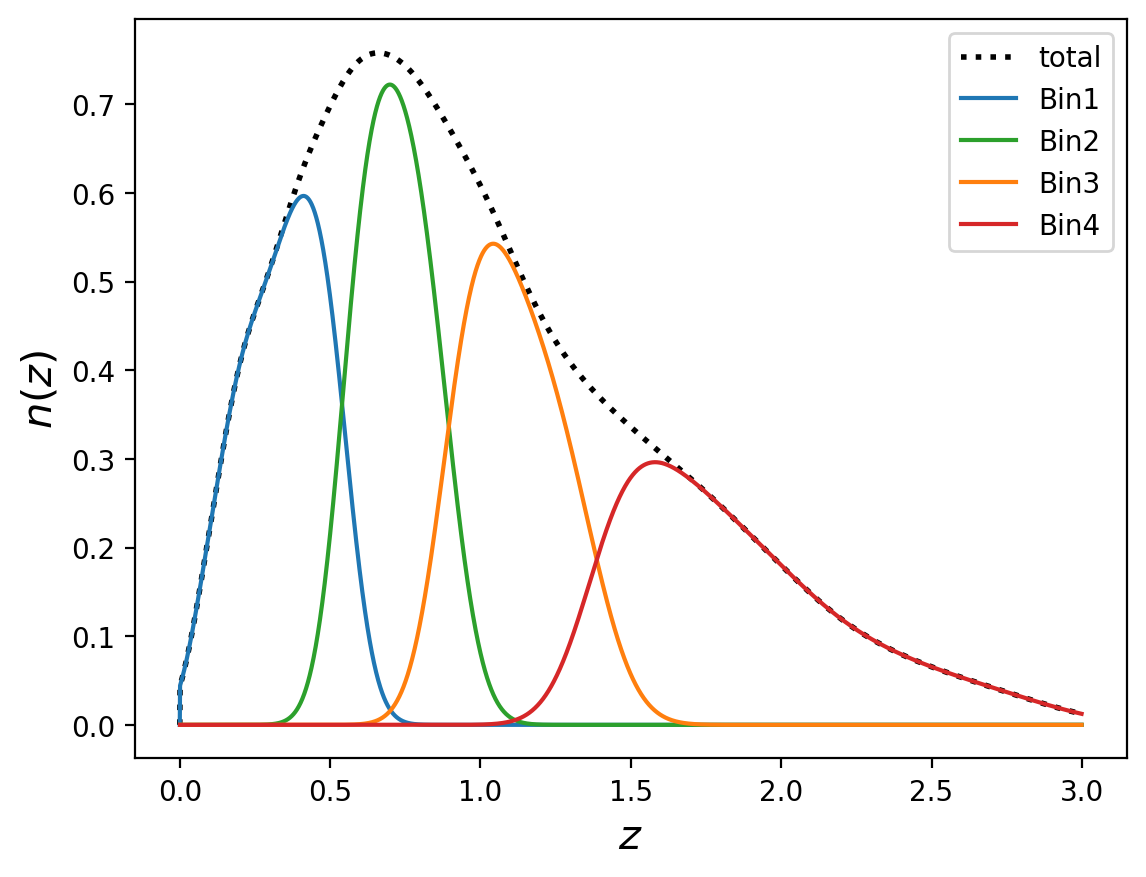}
\caption{The galaxy redshift distributions of the four tomographic bins in the CSST photometric survey derived from the COSMOS catalog \cite{Xiong24}. The dotted curve denotes the total distribution.} 
\label{fig:photoz}
\end{figure}

To investigate the measurement of weak lensing from CSST, we can generate the mock map from simulations and derive the shear correlation function or power spectrum for cosmological analysis. In the left panel of Figure~\ref{fig:WL_map}, we show the CSST mock convergence map in 100 deg$^2$, which is reconstructed from the light-cone given by Jiutian simulations. The convergence map is derived from  the shear field at $0.55<z<0.89$, which is lensed by the matter distribution at $0<z<0.55$, by applying the Kaiser-Squires inversion (KS-inversion) \cite{Kaiser93}. The reconstruction details can be found in \cite{Xiong24}.

In \cite{Xiong24}, by dividing the redshift range from $z=0$ to 3 into four tomographic bins, the mock data of the auto and cross convergence power spectra between different bins were derived. The galaxy photo-$z$ distributions in the four bins are shown in Figure~\ref{fig:photoz}. The galaxy photometric redshift can be well estimated by the SED template fitting methods or mechine learning with the accuracy better than $\sigma_{z0}\simeq0.03$ \cite{Cao18,Zhou21,Zhou22a, Zhou22b,Lu24,Luo24a,Luo24b}. Considering the cosmological parameters, such as the reduced Hubble constant $h$, matter density parameter $\Omega_{\rm m}$, baryon density parameter $\Omega_{\rm b}$, the amplitude parameter of matter fluctuation on a scale of 8~$h^{-1}$Mpc $\sigma_8$, spectral index $n_{\rm s}$, and the equation of stat of dark energy $w$, the intrinsic alignment, i.e. $A_{\rm IA}$ and $\eta_{\rm IA}$, and photo-$z$ shift parameter $\Delta z^i$, photo-$z$ stretch parameter $\sigma_z^i$, and shear calibration parameter $m_i$ in each tomographic bin, they performed the Markov Chain Monte Carlo (MCMC) to jointly constrain the model parameters. 

They found that the CSST weak lensing survey can give $23\%$, $9\%$, and $28\%$ constraint accuracies for $\Omega_{\rm m}$, $\sigma_8$, and $w$ in 100 deg$^2$ survey area, which is comparable to the results from full stage III surveys, such as DES Y3 \cite{Porredon22,Faga24}. If considering the full CSST wide-field survey with 17500 deg$^2$, the constraint accuracy can be improved by nearly one order of magnitude and reach a few percent level for these dark matter and dark energy parameters. This also has been proven by the results derived from the  theoretical estimation given by \cite{Lin22,Lin24}, which gives the relative accuracies $\sim6\%$, $\sim2\%$, and $\sim5\%$ for $\Omega_{\rm m}$, $\sigma_8$, and $w$, respectively. In addition to these cosmological parameters, the CSST weak lensing survey also can effectively constrain the total neutrino mass $\sum m_{\nu}$, and the particle mass and the dark matter fraction of ultra-light axions (ULAs), i.e. $m_a$ and $f_a$. The constraint limits are found to be $\Sigma m_{\nu}\lesssim 0.5$~eV, and $m_a\gtrsim10^{-22.3}$~eV and $f_a\lesssim0.8$ \cite{Lin22,Lin24}.

Besides, the weak lensing survey also can correlate with the two-dimensional (2D) galaxy clustering measurement to obtain three two-point correlation functions, i.e. $3\times2$pt, which can further improve the constraint accuracy.

\subsection{Galaxy clustering}
\subsubsection{2D galaxy clustering}

The measurement of the 2D angular galaxy clustering is another powerful cosmological probe in the CSST photometric survey. The measured two-point angular correlation function of galaxy clustering with angle separation $\theta$ in the $i$th and $j$th tomographic bins are given by
\begin{equation}
w_{ij}(\theta) = \int \frac{{\rm d}\ell\, \ell}{2\pi}\,\tilde{C}_{\rm gg}^{ij}(\ell)\,J_{0}(\ell\theta),
\end{equation}
Here $\tilde{C}_{\rm gg}^{ij}(\ell)$ is the measured angular galaxy power spectrum, which can be expressed as
\begin{equation}
\tilde{C}_{\rm gg}^{ij}(\ell) = {C}_{\rm gg}^{ij}(\ell) + \frac{\delta_{ij}}{\bar{n}_i} + N_{\rm sys}^{ij},
\end{equation}
where $N_{\rm sys}$ is the systematic noise in the survey, which is believed to be quite low ($<10^{-8}$) compared to the shot noise term (the second term on the right) \cite{Gong19}. ${C}_{\rm gg}^{ij}(\ell)$ is the 2D angular galaxy power spectrum, and if assuming flat-sky and Limber approximations it can be expressed as
\begin{equation}
{C}_{\rm gg}^{ij}(\ell) = \int_0^{\chi_{\rm H}} {\rm d}\chi\, \frac{W^i_{\rm g}(\chi)W^j_{\rm g}(\chi)}{r^2(\chi)}\, P_{\rm m}\left( k=\frac{\ell+1/2}{r(\chi)},\chi\right).
\end{equation}
Here $W^i_{\rm g}(\chi)$ is the weighting function of the 2D galaxy clustering, and it is given by
\begin{equation}
W^i_{\rm g}(\chi) = b_{\rm g}^i\, n_i(\chi),
\end{equation}
where $b_{\rm g}^i$ is the galaxy bias in the $i$th tomographic bin, which can be set as a free parameter. Since the theoretical model is not accurate enough at small scales in the non-linear regime, only the data at large scales ($k<0.3\ {\rm Mpc}^{-1}h$) are retained in the angular galaxy power spectrum analysis.

In the right panel of Figure~\ref{fig:WL_map}, we show the CSST galaxy density mock map at $0<z<0.55$ in 100 deg$^2$ survey area derived from Jiutian simulations \cite{Xiong24}. The 2D galaxy auto and cross power spectra between the four tomographic bins in $0<z<3$, as the CSST weak lensing case discussed above, can be estimated. After applying the MCMC method, we can jointly constrain the cosmological and systematical parameters, including the galaxy bias $b_{\rm g}^i$ in each photo-$z$ bin. 

Since only the data at large scales can be safely used in the angular galaxy power spectrum for cosmological analysis, the constraint power of the 2D galaxy clustering is not as strong as the weak lensing. For example, \cite{Lin22} and \cite{Lin24} find that the constraint accuracies of $\Omega_{\rm m}$, $\sigma_8$, and $w$ are about $10\%$, $3\%$, and $13\%$, respectively, and the constraint limits of $\sum m_{\nu}$, $m_a$, and $f_a$ are $\lesssim 0.7$~eV, $\gtrsim10^{-23.4}$~eV, and $\lesssim0.9$, respectively.

However, if considering the cross correlation with the weak lensing survey and cooperating the shear, galaxy angular, and galaxy-galaxy lensing correlation functions or power spectra, i.e. the $3\times2$pt, the constraint accuracy can be significantly improved. The galaxy-galaxy lensing cross correlation function with angle separation $\theta$ in the $i$th and $j$th tomographic bins can be expressed as
\begin{equation}
\gamma_{\rm t}^{ij}(\theta) = \int \frac{{\rm d}\ell\, \ell}{2\pi}\,{C}_{\gamma \rm g}^{ij}(\ell)\,J_{2}(\ell\theta),
\end{equation}
where $J_{2}(\ell\theta)$ is the second order Bessel function of the first kind. By including the intrinsic alignment effect, the galaxy-galaxy lensing power spectrum can be written as
\begin{equation}
 {C}_{\gamma \rm g}^{ij}(\ell) =  {C}_{\kappa \rm g}^{ij}(\ell) +  {C}_{\rm Ig}^{ij}(\ell),
\end{equation}
where ${C}_{\kappa \rm g}^{ij}(\ell)$ and ${C}_{\rm Ig}^{ij}(\ell)$ are given by
\begin{equation}
{C}_{\kappa \rm g}^{ij}(\ell) =  \int_0^{\chi_{\rm H}} {\rm d}\chi\, \frac{W^i_{\kappa}(\chi)W^j_{\rm g}(\chi)}{r^2(\chi)}\, P_{\rm m}\left(\frac{\ell+1/2}{r(\chi)},\chi\right),
\end{equation}
and
\begin{equation}
{C}_{\rm Ig}^{ij}(\ell) =  \int_0^{\chi_{\rm H}} {\rm d}\chi\, \frac{n_i(\chi)F_i(\chi)W^j_{\rm g}(\chi)}{r^2(\chi)}\, P_{\rm m}\left(\frac{\ell+1/2}{r(\chi)},\chi\right).
\end{equation}

As shown in Figure~\ref{fig:WL_map}, the convergence map at $0.55<z<0.89$ can well match the galaxy density map at $0<z<0.55$, which indicates there is a strong correlation between the two fields, i.e. a large galaxy-galaxy lensing power spectrum. By utilizing the CSST $3\times2$pt, we can find that the constraint accuracy of $\Omega_{\rm m}$, $\sigma_8$, and $w$ are $\sim2\%$, $0.5\%$, and $3\%$, and the constraint limits of $\sum m_{\nu}$, $m_a$, and $f_a$ are $\lesssim 0.4$~eV, $\gtrsim10^{-21.9}$~eV, and $\lesssim0.7$, respectively \cite{Lin22,Lin24}. Note that we can use the same tomographic bins for the weak lensing and 2D galaxy clustering surveys, which can help to effectively constraint the photo-$z$ systematics, i.e. $\Delta z$ and $\sigma_z$ \cite{Lin22,Lin24}.

\subsubsection{3D galaxy clustering}

In addition to the multi-band imaging photometric survey, the CSST also can perform the wide-field spectroscopic galaxy survey with slitless gratings. In a real spectroscopic survey, we actually measure the distribution of galaxies in redshift space. The redshift-space galaxy power spectrum in the $i$th spectroscopic redshift (spec-$z$) bin can be expanded in Legendre polynomials, which is given by
\begin{equation}
P_{\rm g}^i(k,\mu) = \sum_{\ell} \tilde{P}_{\ell}^{{\rm g},i}(k)\, \mathcal{L}_{\ell}(\mu),
\end{equation}
where $\mu=k_{\parallel}/k$ is the cosine of the angle between $k$ and the line of sight, and $\mathcal{L}_{\ell}(\mu)$ is the Legendre polynomials. Note that only the orders with $\ell=(0, 2, 4)$ are not vanishing in the linear regime. $\tilde{P}_{\ell}^{{\rm g},i}(k)$ is the total galaxy multipole power spectrum in the $i$th spec-$z$ bin, that we have
\begin{equation}
\tilde{P}_{\ell}^{{\rm g},i}(k) = P_{\ell}^{{\rm g},i}(k) + \frac{1}{\bar{n}_{\rm g}^i},
\label{eq:Plg}
\end{equation}
where ${\bar{n}_{\rm g}^i}$ is the average galaxy number density in the $i$th spec-$z$ bin. $P_{\ell}^{\rm g}(k)$ is the galaxy multipole power spectrum, and if considering the Alcock-Paczynski (AP) effect \cite{Alcock79}, it is given by
\begin{equation}
P_{\ell}^{\rm g}(k) = \frac{2\ell+1}{2q^2_{\perp}q_{\parallel}} \int_{-1}^1 {\rm d}\mu\, P_{\rm g}(k',\mu') \, \mathcal{L}_{\ell}(\mu).
\end{equation}
Here $q_{\perp}=D_{\rm M}(z)/D^{\rm fid}_{\rm M}(z)$ and $q_{\parallel}=H^{\rm fid}(z)/H(z)$ are the AP scaling factors,  where $D_{\rm M}(z)$ is the comoving angular diameter diatance, and ``fid" means the fiducial cosmology. $k'=\sqrt{k^{'2}_{\perp}+k^{'2}_{\parallel}}$ and $\mu'=k'_{\parallel}/k'$, where $k'_{\perp}=k_{\perp}/q_{\perp}$ and $k'_{\parallel}=k_{\parallel}/q_{\parallel}$. If assuming the matter field is completely correlated with the velocity field, i.e. there is no peculiar velosity bias, the apparent redshift-space galaxy power spectrum can be expressed as
\begin{equation}
P_{\rm g}(k',\mu') = P_{\rm g}(k')\, (1+\beta\mu'^2)^2\,\mathcal{D}(k',\mu').
\end{equation}
Here $P_{\rm g}(k')=b_{\rm g}^2\,P_{\rm m}(k')$ is the apparent real-space galaxy power spectrum, $\beta=f/b_{\rm g}$ where $f$ is the growth rate, and $\mathcal{D}(k',\mu')$ is the damping term which mainly affect the small scales. It can be estimated by
\begin{equation}
\mathcal{D}(k',\mu') = e^{-(k'\mu'\sigma_{\rm D})^2},
\end{equation}
where $\sigma_{\rm D}^2=\sigma_{\nu}^2+\sigma_{R}^2$, which account for the galaxy velocity dispersion and the smearing effect due to the spectral resolution of the spectroscopic survey, respectively \cite{Gong19}.

\begin{figure}[H]
\centering
\includegraphics[scale=0.55]{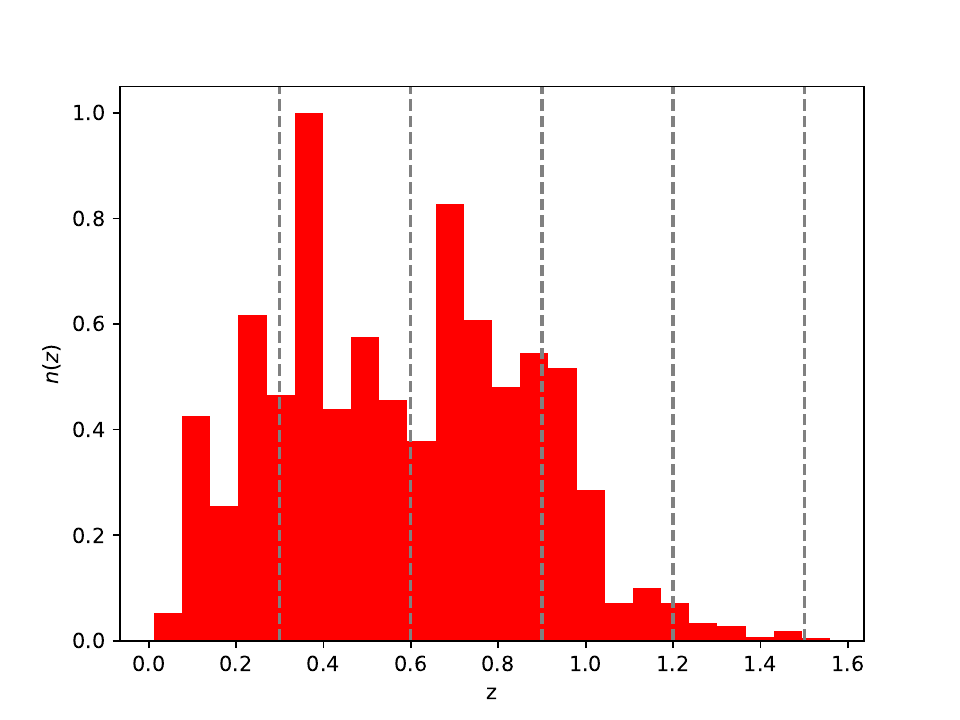}
\caption{The galaxy redshift distribution of the CSST spectroscopic survey derived from the zCOSMOS catalog \cite{Miao23}. The vertical dashed lines denote the five spec-$z$ bins.} 
\label{fig:specz}
\end{figure}

In Eq.~(\ref{eq:Plg}), the shot noise term $1/\bar{n}_{\rm g}$ or the galaxy number density $\bar{n}_{\rm g}$ is crucial for the galaxy spectroscopic survey, which can directly affect the error of the galaxy power spectrum. In the CSST galaxy spectroscopic survey, the total surface galaxy number density can reach $\sim2-3$ arcmin$^{-2}$, which gives $\bar{n}_{\rm g}>10^{-2}\ {\rm Mpc}^{-3}h^3$ at $0<z<0.6$ and $\bar{n}_{\rm g}>10^{-3}\ {\rm Mpc}^{-3}h^3$ at $0.6<z<1.2$ \cite{Gong19}. This number density is much higher than the Stage III galaxy surveys, e.g. eBOSS \cite{Dawson16}. The estimated galaxy redshift distribution of the CSST spectroscopic survey is shown in Figure~\ref{fig:specz} \cite{Miao23}. We can see that there is a peak at $z\sim0.3$, and the distribution can extend to $z>1.5$.

We should note that, since the slitless gratings are adopted in the CSST spectroscopic survey, it is quite challenging for data processing to derive accurate spec-$z$ for galaxies. If the spec-$z$ accuracy cannot achieve the required accuracy with $\sigma_{z0}\simeq0.002$, the measurement of the galaxy power spectrum would be affected significantly, especially at small scales. If we only take the data in the linear regime at $k\lesssim0.1-0.2\ {\rm Mpc}^{-1}h$, this effect can be reduced, and the required spec-$z$ accuracy may be released to be as large as $\sigma_{z0}\simeq0.005$. The galaxies with larger spec-$z$ uncertainties need to be discarded in the analysis, which will suppress $\bar{n}_{\rm g}$ in a spec-$z$ bin and lead to larger shot noise. As shown in \cite{Zhou21,Zhou24}, the CSST galaxy spec-$z$ accuracy can reach $\sim0.001$, if using machine learning with suitable training datasets.

In \cite{Gong19}, the constraint result indicates that the CSST 3D galaxy clustering measurements can provide comparable constraint strength on the cosmological parameters, e.g. $\Omega_{\rm m}$, $\sigma_8$, and $w$, to that from the weak lensing survey. Besides, as indicated in \cite{Chen22}, the CSST galaxy spectroscopic survey can effectively constrain the modified gravity models, e.g. Brans-Dicke theory. They found that the the constraint strength is comparable to the results from the joint dataset from the current cosmic microwave background (CMB), BAO and SNe Ia observations.

\subsection{Galaxy cluster abundance}

The galaxy cluster abundance or number counts is also a useful cosmological probe. Galaxy clusters are large gravitationally bound systems in the Universe, which represent the most dense regions with mass $M=10^{14}-10^{15}\ M_{\odot}$ containing hundreds to thousands of galaxies. The number of galaxy clusters at a given redshift per steradian is given by
\begin{equation}
\frac{{\rm d}^2N}{{\rm d}\Omega{\rm d}z} = \frac{{\rm d}n(M,z)}{{\rm d}M}\frac{{\rm d}V}{{\rm d}\Omega{\rm d}z}{\rm d}M,
\end{equation}
where ${\rm d}n(M,z)/{\rm d}M$ is the halo mass function \cite{Sheth99,Tinker08}. Then the expected value of galaxy cluster number counts in the $i$th mass bin and $m$th redshift bin is
\begin{equation}
N_{mi} = \Omega_{\rm sky} \int_{\Delta z_m} {\rm d}z\, \frac{{\rm d}V}{{\rm d}\Omega{\rm d}z} \int_{\Delta M_i} {\rm d}M\, \frac{{\rm d}n(M,z)}{{\rm d}M}.
\end{equation}
Here $\Omega_{\rm sky}=A_{\rm s}(\pi/180)^2$ is the sky coverage, where $A_{\rm s}$ is the survey area in square degree, and ${\rm d}V/{\rm d}\Omega{\rm d}z=r^2(z)\,c/H(z)$, where $r(z)$ and $H(z)$ are the comoving distance and Hubble parameter at $z$, respectively.

The challenge in galaxy cluster number counts is to precisely determine the redshift and mass of a galaxy cluster, which needs both photometric and spectroscopic measurements in optical galaxy surveys. Since CSST can simultaneously perform the photometric and spectroscopic surveys, it has significant advantages in the observation of galaxy cluster number counts. For instance, the cluster redshift and members can be determined by its spectroscopic survey, and the cluster mass can be derived from its gravitational lensing observation and the richness obtained in its photometric survey.

\begin{figure}[H]
\centering
\includegraphics[scale=0.55]{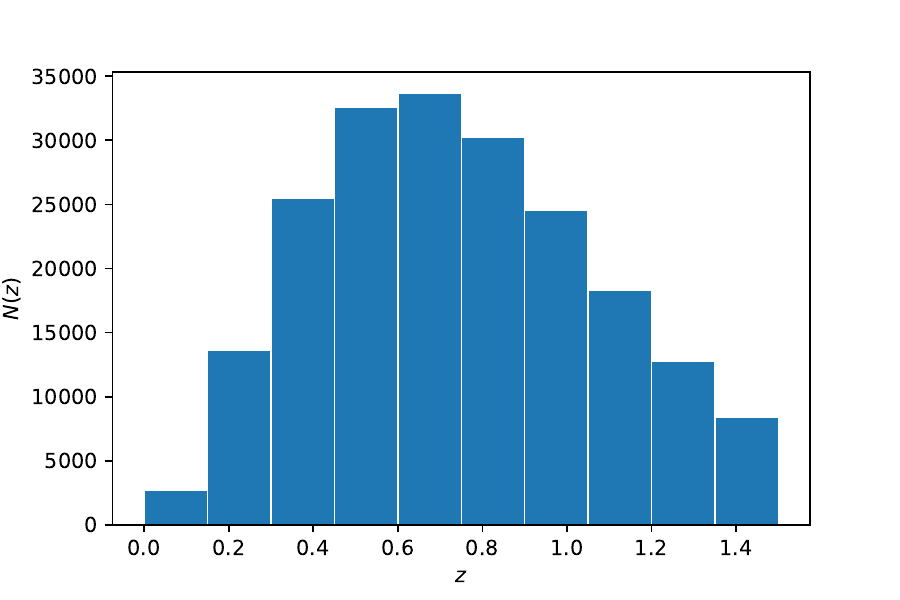}
\caption{The predicted galaxy cluster redshift distribution identified by the CSST photometric and spectroscopic surveys \cite{Miao23}.} 
\label{fig:cluster}
\end{figure}

In Figure~\ref{fig:cluster}, the expected galaxy cluster redshift distribution from CSST has been shown \cite{Miao23}. In order to ensure accurate measurement of the redshift and mass of galaxy clusters, only the redshift range $0<z<1.5$ is considered, which can be covered by both the CSST photometric and spectroscopic surveys. In \cite{Miao23}, they found that CSST can detect about 170,000 clusters within the mass range $10^{14}-10^{16}\ h^{-1}M_{\odot}$, and it could be as large as about 400,000 if including the clusters with lower mass \cite{Zhang23}. The constraint accuracies of $\Omega_{\rm m}$, $\sigma_8$, and $w$ are found to be $\sim$5\%, 1\%, and 13\%, respectively, which has strong constraint on $\Omega_{\rm m}$ and $\sigma_8$. This is even more stringent than the weak lensing and 3D galaxy clustering surveys as discussed in Section 2.1 and 2.2 \cite{Gong19,Miao23,Lin22,Lin24}.

We also notice that, in addition to galaxy cluster number counts, CSST also can use other probes related to galaxy clusters, such as the two-point correlation function and strong lensing, to constrain the cosmological parameters or models. All these measurements can further improve the constraint power of galaxy clusters.

\subsection{Cosmic void}

\begin{figure}[H]
\centering
\includegraphics[scale=0.29]{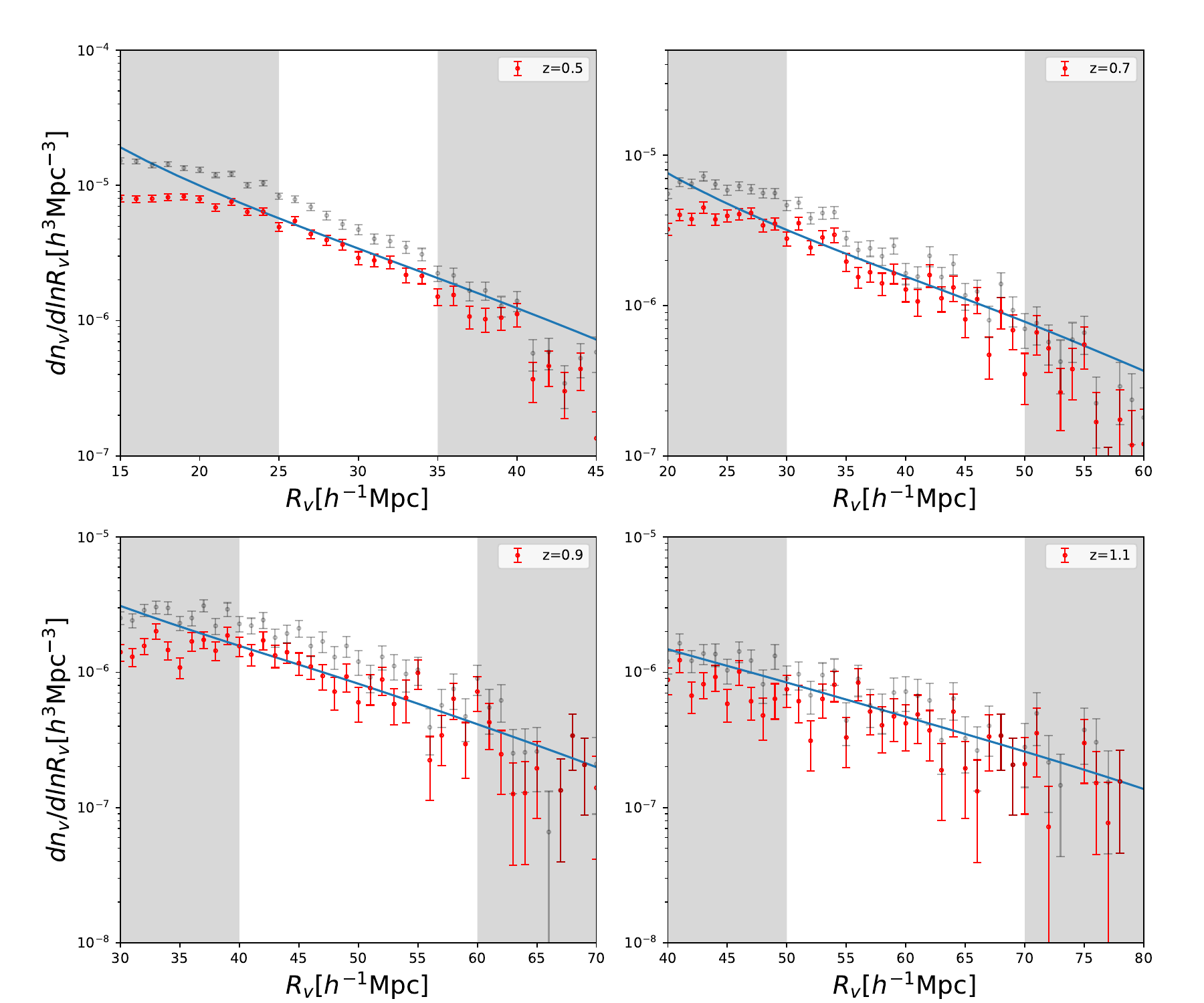}
\caption{The predicated VSFs measured from Jiutian simulations with box size $1\ h^{-1}{\rm Gpc}$ for the CSST spectroscopic survey around $z=0.5$, 0.7, 0.9 and 1.1 \cite{Song24a}. The red and gray data points denote the VSFs with $\epsilon_{\rm v}<0.15$ and all voids, respectively. The blue curves are the best-fit theoretical model. The gray regions represent the scales discarded in the analysis.} 
\label{fig:VSF}
\end{figure}

Besides the measurements in dense regions of the Universe, the probes of low-density regions, e.g. cosmic voids, also contain rich cosmological information. Since cosmic voids have the characteristics of large volume, low density, and linear evolution, they are particularly suitable for studying the growth of the LSS, property of dark energy, modified gravity theories, etc. For example, the void size function (VSF) is a widely used probe in cosmological study, which denotes the number density of voids as a function of size at a given redshift, and it can be expressed as
\begin{equation}
\frac{{\rm d}n_{\rm v}}{{\rm d\,ln}R_{\rm v}}=\frac{3}{4\pi R_{\rm v}^3}\,F(\nu,\delta_{\rm v},\delta_{\rm c})\, \frac{d\nu}{{\rm d\,ln}R_{\rm L}}.\label{eq:vsf}
\end{equation}
Here $R_{\rm v}$ and $R_{\rm L}$ are the void Eulerian and Lagrangian sizes, respectively, and $\delta_{\rm v}=|\delta_{\rm v}|/\sigma_{\rm M}(z)$, where $\delta_{\rm v}$ is the linear underdensity threshold for void formation and $\sigma_{\rm M}(z)$ is the root-mean-square density fluctuations. $F(\nu,\delta_{\rm v},\delta_{\rm c})$ is the first-crossing distribution, which denotes the probability of a random trajectory crossing the barrier $\delta_{\rm v}$ for the first time at $\nu$ without crossing $\delta_{\rm c}=1.686$ for $\nu'>\nu$. More details can be found in \cite{Song24a}. 

In Figure~\ref{fig:VSF}, the predicated VSFs measured by the CSST spectroscopic survey around $z=0.5$, 0.7, 0.9 and 1.1, using Jiutian simulations with box size $1\ h^{-1}{\rm Gpc}$, have been shown \cite{Song24a}. They identified voids from the mock CSST galaxy catalog using the Voronoi tessellation and watershed algorithm without assuming any void shape. They found that the CSST VSF measurement can constrain the cosmological parameters, e.g. $\Omega_{\rm m}$ and $w$, to a few percent level, and the void parameter $\delta_{\rm v}$ also can be jointly constrained in high precision. Besides, the degeneracy direction of $\Omega_{\rm m}$ and $\sigma_8$ is almost orthogonal to that from galaxy clustering measurements, which indicates the VSF can be a good complementary probe to the measurements of overdense regions \cite{Contarini22,Pelliciari23,Song25a}. 

In addition, we can also use void number counts (VNC) to extract the cosmological information. The VNC can be estimated by integrating the VSF over the void radius $R_{\rm v}$ in a survey volume $V_{\rm s}$ at a given redshift, and we have
\begin{equation}
N_{\rm v}(z) = V_{\rm s} \int_{\Delta R_{\rm v}} \frac{{\rm d}n(R_{\rm v},z)}{{\rm d}R_{\rm v}}{\rm d}R_{\rm v}.
\end{equation}
In \cite{Song24c}, they found the constraint power of the VNC is comparable to that from the VSF, although the VNC is the integral of the VSF. This is because that the VNC is not sensitive to void shape, and more voids can be included in the VNC analysis than the VSF by applying simpler selection criteria.

Similar to galaxy clusters, we can employ other probes related to voids, such as the 2D or 3D void auto and void-galaxy cross correlation function or power spectra \cite{Song25b,Song24d}, void gravitational lensing, etc. It is expected that the cosmic voids will be more and more important in the next-generation surveys, which explore huge sky volume with high resolution and sensitivity.

\section{Cosmic distance measurements}\label{sec:3}

The CSST also can accurately measure the cosmic distance at different redshifts or the expansion history of the Universe. In this section, we discuss the the SN Ia and BAO measurements, which are known as the standard candle and ruler, in the CSST surveys.

\subsection{Type Ia supernovae}

The CSST plans to implement a 9 deg$^2$ UDF survey in the first two years with a  single exposure time 250s and 60 exposures. This means that it will observe a patch of sky in the UDF averagely every $\sim$12 days, which is close to the best cadence found by \cite{Li23} for the CSST SN~Ia survey, and the magnitude limit of a single exposure in the photometric observation can reach $i=25.9$ (see Table~\ref{tab:CSST_para}). Therefore, CSST has great potential in SN Ia detection, and can measure the light curve of a SN~Ia precisely, considering its survey cadence and depth. 

After fitting the SN~Ia light curve, the redshift $z$, explosion time $t_0$, and other three parameters related to the distance modulus  can be derived. Then the observed distance modulus of a SN~Ia is given by
\begin{equation}
\mu = m_{B} + \alpha x_1 - \beta c - M_0,
\end{equation}
where $m_B$ is the $B$-band apparent magnitude, $x_1$ is the time-dependent variation, and $c$ is the color parameter that is independent of time, and $\alpha$, $\beta$, and $M_0$ are nuisance parameters which can be set as free parameters in the fitting process.
On the other hand, the theoretical distance modulus can be calculated by
\begin{equation}
\mu_{\rm th}(z) = 5\,{\rm log}_{10}\,d_{\rm L}(z) + 25,
\end{equation}
where $d_{\rm L}(z)$ is the luminosity distance at $z$, which is determined by the cosmological model. 

\begin{figure}[H]
\centering
\includegraphics[scale=0.16]{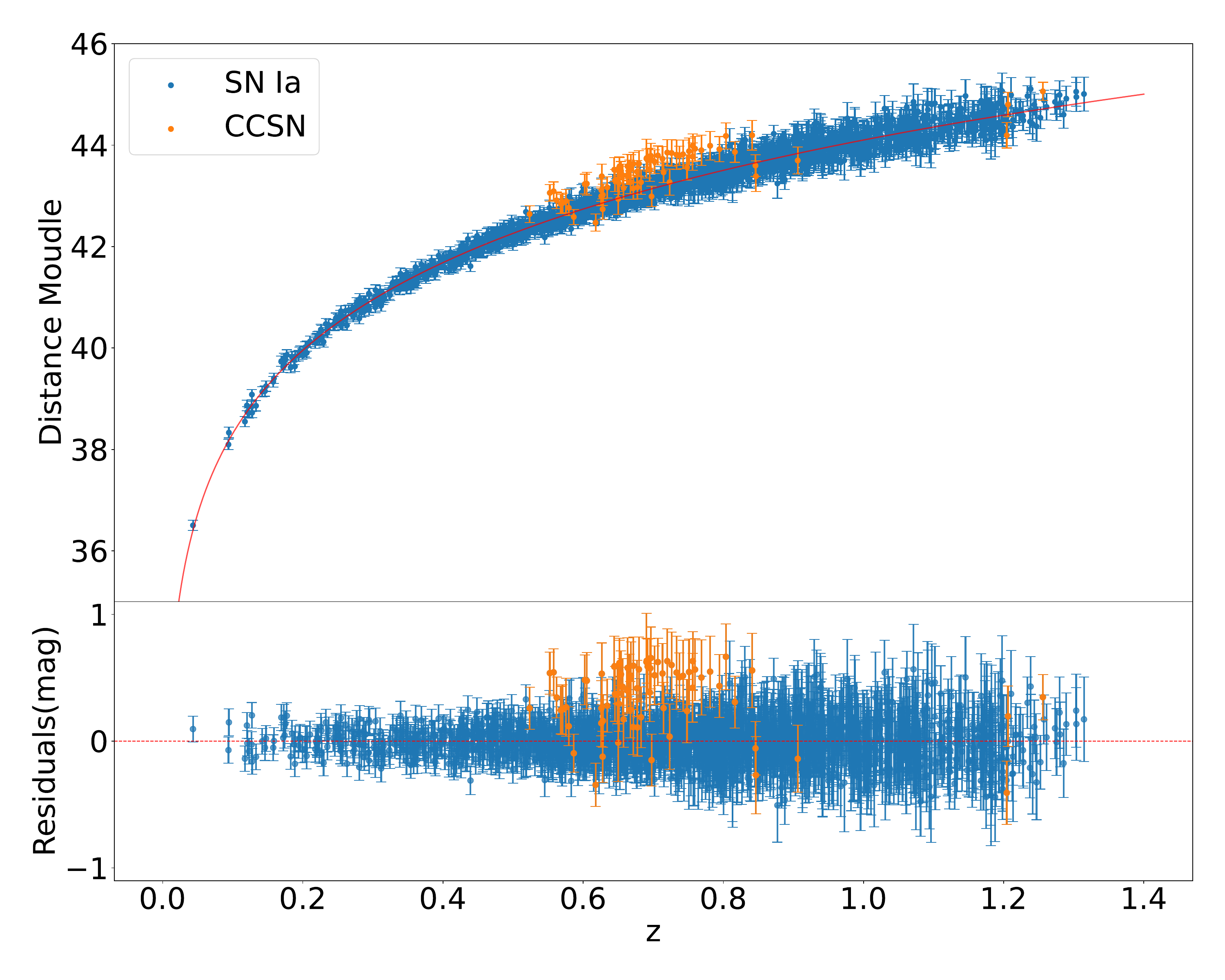}
\caption{The Hubble diagram for the SN Ia photometric mock sample derived from the CSST UDF survey \cite{Wang24}. This sample has been calibrated to reduce the contamination of CCSNe. The blue and orange data points are for SNe~Ia and CCSNe, respectively. The red curve is the best-fit theoretical model.} 
\label{fig:HD}
\end{figure}

In \cite{Wang24}, they generated the light-curve mock data for SNe Ia and different types of core-collapse SNe (CCSNe) which can be detected by the CSST-UDF photometric survey. After selecting high-quality data, fitting the light curves and applying contamination reduction, the SN~Ia light-curve parameters are derived, and CCSNe as contamination are identified. Finally, about $\sim 2000$ SNe~Ia and $\sim70$ CCSNe are left for the cosmological analysis. We note that, if considering the CSST spectroscopic observation in the UDF, the contamination of CCSNe can be further reduced by analyzing SN spectra. In Figure~\ref{fig:HD}, the Hubble diagram as a function of redshift for the CSST SN~Ia mock sample has been shown \cite{Wang24}. The cosmological and SN~Ia nuisance parameters are jointly fitted using this sample. They found that the residual CCSNe would not affect the results, and the constraint accuracies on $\Omega_{\rm m}$ and $w$ are about two times better than the current SN surveys \cite{Brout22,DES24}.

Besides the CSST-UDF survey, the CSST wide-field survey also can be used to detect a large number of SNe in ten years, although it is not dedicated to time-domain studies. As shown in \cite{Liu24}, about 5 million SNe of various types can be detected by CSST for follow-up observations by time-domain telescopes. This will greatly promote future SN studies, which can be helpful to measure the Hubble constant $H_0$ and paramter $H(z)$, and probe the expansion history of the Universe with extreme accuracy.

\subsection{Baryonic acoustic oscillations}

Another powerful probe for measuring cosmic distance or expansion history is the BAO, which can be seen as a standard ruler. The BAO is directly related to the comoving angular diameter distance $D_{\rm M}$ in transverse direction, the Hubble equivalent distance $D_{\rm H}\equiv c/H(z)$ in radial direction, and the comoving sound horizon at the drag epoch or recombination $r_{\rm d}$. We can extract the BAO signal or distance information by fitting the galaxy two-point correlation function or power spectrum. 

In practice, we usually constrain two BAO scaling factors $\alpha_{\perp}$ and $\alpha_{\parallel}$, which contain the cosmic distance information and can be used to analyze the cosmology. When considering the AP effect, the two scaling factors at redshift $z$ can be expressed as
\begin{equation}
\alpha_{\perp}(z) = \frac{D_{\rm M}(z)\,r_{\rm d}^{\rm fid}}{D_{\rm M}^{\rm fid}(z)\,r_{\rm d}},\ \ \ {\rm and}\ \ \ \alpha_{\parallel}(z) = \frac{D_{\rm H}(z)\,r_{\rm d}^{\rm fid}}{D_{\rm H}^{\rm fid}(z)\,r_{\rm d}},
\end{equation}
where $r_{\rm d}$ is given by
\begin{align}
r_{\rm d}  =& \int_{\infty}^{z_{\rm d}} \frac{c_{\rm s}(z)}{H(z)}{\rm d}z, \nonumber\\
 \simeq& \frac{147.05}{\rm Mpc}\left( \frac{\Omega_{\rm m}h^2}{0.1432}\right)^{-0.23} \left( \frac{N_{\rm eff}}{3.04}\right)^{-0.1} \left( \frac{\Omega_{\rm b}h^2}{0.02235}\right)^{-0.13}.
\label{eq:2}
\end{align}
Here $c_{\rm s}$ is the sound speed and $N_{\rm eff}$ is the effective number of neutrino species. After extracting the two BAO scaling factors, the distance observable can be estimated by
\begin{equation}
\frac{D_{\rm M}(z)}{r_{\rm d}} = \alpha_{\perp} \frac{D_{\rm M}^{\rm fid}(z)}{r_{\rm d}^{\rm fid}},\ \ \ {\rm and}\ \ \ \frac{D_{\rm H}(z)}{r_{\rm d}} = \alpha_{\parallel} \frac{D_{\rm H}^{\rm fid}(z)}{r_{\rm d}^{\rm fid}}.
\end{equation}

Since the CSST wide-field galaxy spectroscopic survey can explore huge space volume with 17,500 deg$^2$ survey area at $0<z<1.5$, and observe more than one hundred million galaxy spectra, as shown in Table~\ref{tab:CSST_para} and discussed in Section~2.2.2, it can precisely measure the BAO signal. Besides, the CSST is also expected to obtain more than four million AGN spectra at $0<z<5$ \cite{Miao24}, which also can be used to measure the BAO covering a large redshift range. 

\begin{figure}[H]
\centering
\includegraphics[scale=0.3]{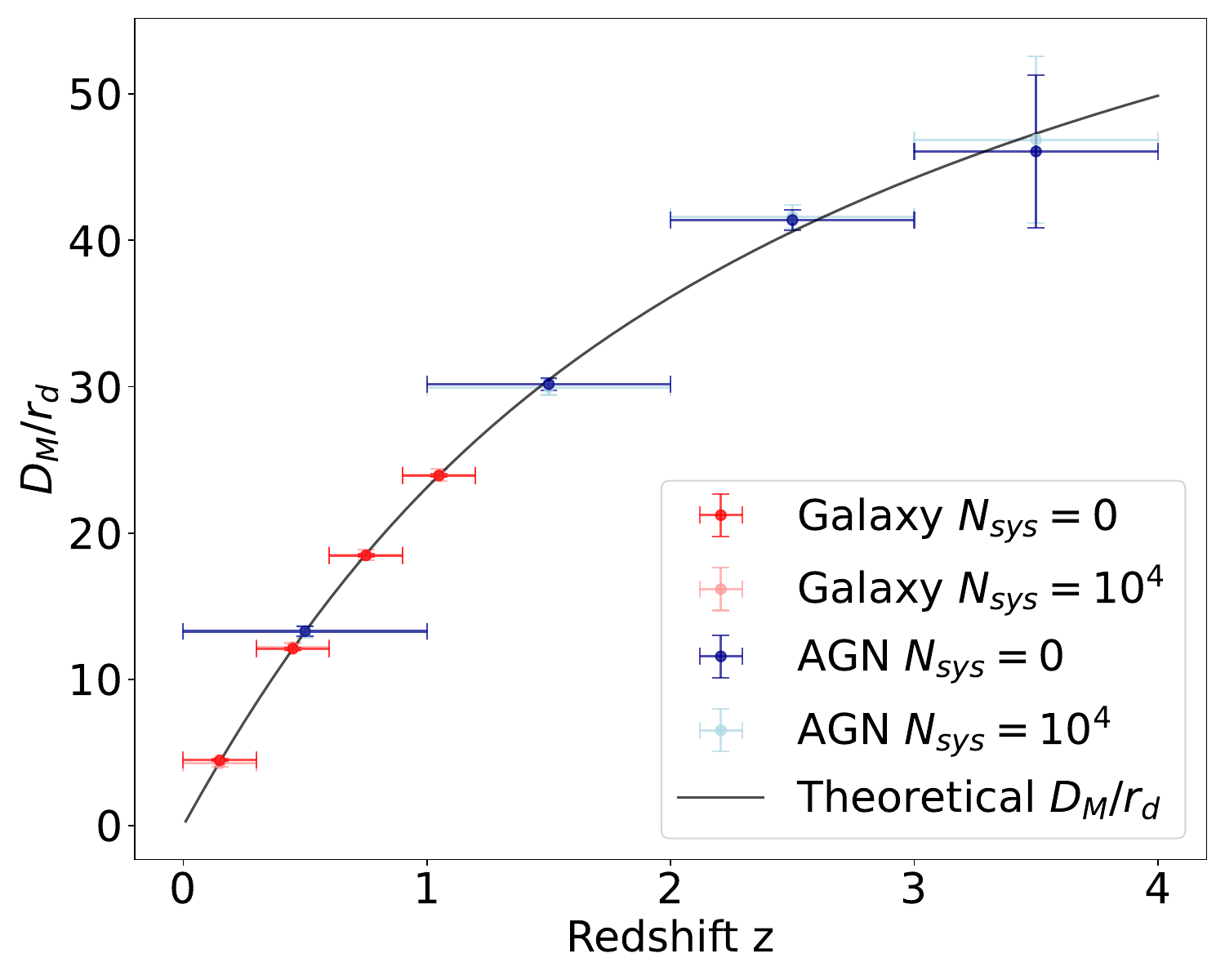}
\includegraphics[scale=0.3]{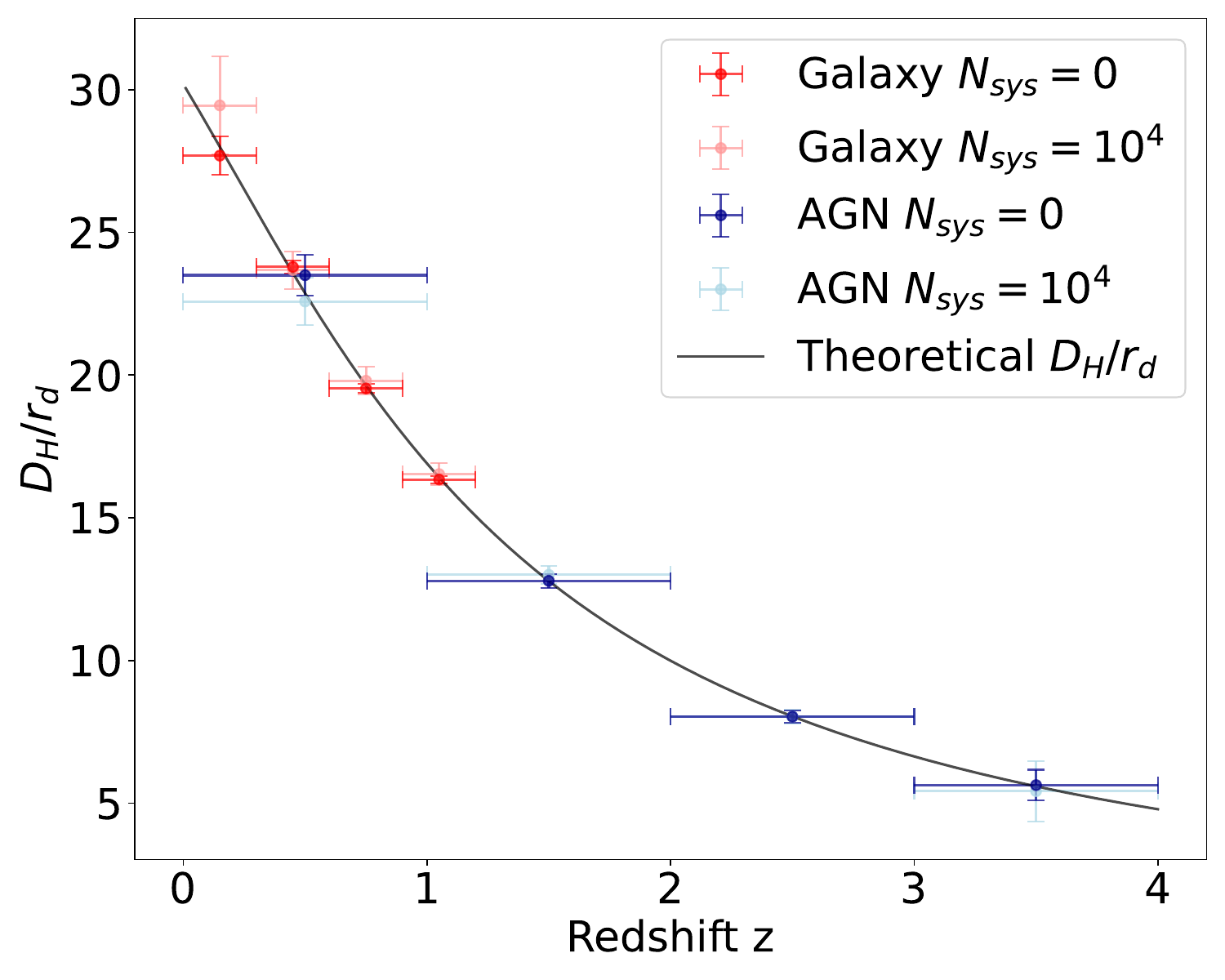}
\caption{The $D_{\rm M}/r_{\rm d}$ and $D_{\rm H}/r_{\rm d}$ as a function of redshift derived from the CSST BAO measurements for galaxies (red) and AGNs (blue) assuming different levels of systematics \cite{Yan24,Miao24}. The curves indicate the best-fit theoretical model.} 
\label{fig:D_MH}
\end{figure}

In Figure~\ref{fig:D_MH}, we show the prediction of the measured $D_{\rm M}/r_{\rm d}$ and $D_{\rm H}/r_{\rm d}$ as a function of redshift from the CSST spectroscopic survey \cite{Yan24,Miao24}. In \cite{Miao24}, by utilizing the reconstruction technique, they found that the CSST can conduct accurate BAO measurements with precisions higher than $\sim$1\% for the galaxy survey at $0.3<z<1.2$ and $\sim$3\% for the AGN survey at $z<3$. This means that the CSST BAO survey can constrain $\Omega_{\rm m}$ and $w$ to about 4\% and 9\%, respectively.

Note that the accuracy of spec-$z$ is important for the BAO measurements, especially for $\alpha_{\parallel}$ in the radial direction. Since CSST adopts slitless gratings to perform the spectroscopic survey with $R\gtrsim200$, the spec-$z$ accuracy may not be higher than $\sigma_{z0}=0.002$. As discussed in \cite{Miao24} and \cite{Shi25}, while $\alpha_{\perp}$ is not sensitive to $\sigma_{z0}$, only when $\sigma_{z0}<0.005$, $\alpha_{\parallel}$ can be well measured.

Furthermore, using the CSST SN Ia and BAO joint dataset, the $f(R)$ modified gravity theory can be precisely constrained \cite{Yan24}. In \cite{Yan24}, they adopted the above CSST SN Ia and BAO mock data to investigate the constraint on three models of $f(R)$ theory, i.e. the Hu-Sawicki, Starobinsky and ArcTanh models. They found that the constraint accuracy of $f(R)$ theory from the CSST SN Ia+BAO data is comparable to or even better than the result given by the combination of all of the current SN Ia and BAO observations.

\section{Discussion and conclusions}\label{sec:5}
In this work, we review the main cosmological probes in the CSST photometric and spectroscopic surveys and the constraint power on dark energy, dark matter, and modified gravity theories. CSST can precisely measure matter distribution of the Universe by probing weak gravitational lensing, 2D and 3D galaxy clustering, galaxy cluster abundance and cosmic void, and detect the cosmic expansion history by SN~Ia, and BAO observations. By implementing these measurements, CSST will perform joint constraints on cosmological parameters and obtain extremely precise results by effectively breaking the parameter degeneracies.

We find that CSST can measure the equation of state of dark energy with an accuracy better than 5\% and even close to $\sim$1\%, which can determine if dark energy is the cosmological constant or has dynamical evolution as some scalar field or other models. The property of dark matter, e.g. cold or warm, can be clearly distinguished, and the dark matter particle mass (e.g. axion) will be precisely measured, which will provide important reference for detecting dark matter particles in the laboratory. Besides, the theory of gravity can be tested in high precision, and modified gravity models, e.g. Brans-Dicke theory and $f(R)$ theory, could be stringently constrained.

On the other side, it is worth noting that combining probes is not a trivial task, since different probes may suffer different kinds of systematics, and their results may have tension. This brings a challenge to the joint cosmological constrints of CSST and other Stage~IV surveys. However, by adopting more advanced techniques of the data processing and analysis, e.g. mechine learning, we expect that this issue can be well solved. Besides, many parameters in the analyses mentioned above are adjustable (e.g., the number of redshift bins and the range of each bin, the scale range of galaxy clustering analysis, the definition of cosmic voids, etc.). Adjusting these options might further enhance the constraint power, and improve the fitting accuracy.

In addition to the probes discussed above, CSST also can perform other cosmological probes, e.g. the strong gravitational lensing   \cite{Cao24,Wu25}, 2D BAO \cite{Ding24}, AP effect \cite{Xiao23}, cosmic optical background \cite{Cao22}, galaxy-ellipticity correlation \cite{Xu23}, etc. Moreover, CSST also has strong synergy with other telescopes, such as other Stage IV survey telescopes like $Euclid$ \cite{Liu23}, future CMB experiments \cite{Wang23}, time-domain telescopes for detecting transient sources \cite{Liu24}, the radio telescopes like FAST \cite{Deng22} and MeerKAT \cite{Jiang23} for extracting 21cm signal, and future gravitational wave detectors \cite{Song24}. All these indicate that CSST can greatly promote the relevant cosmological studies in the future, and new physics and opportunities related to new discoveries and theories about dark energy, dark matter, and modified gravity are expected to emerge in abundance.

%%%%%%%%%%%%%%%%%%%%%%%%%%%%%%%%%%%%%%%%%%%%%%%%%%%%%%%
%%% Acknowledgements. 
%%%%%%%%%%%%%%%%%%%%%%%%%%%%%%%%%%%%%%%%%%%%%%%%%%%%%%%
\Acknowledgements{YG acknowledges the support of National Key R\&D Program of China grant Nos. 2022YFF0503404, 2020SKA0110402, and the CAS Project for Young Scientists in Basic Research (No. YSBR-092). XLC acknowledges the support of the National Natural Science Foundation of China through Grant No. 12361141814, and the Chinese Academy of Science grants ZDKYYQ20200008. YPJ acknowledges the support of NSFC 12133006 and National Key R\&D Program of China grant Nos. 2023YFA1607800 and 2023YFA1607801. This work is also supported by science research grants from the China Manned Space Project with grant nos. CMS-CSST- 2021-B01, CMS-CSST-2021-A01, and CMS-CSST-2021-A03.}

%%%%%%%%%%%%%%%%%%%%%%%%%%%%%%%%%%%%%%%%%%%%%%%%%%%%%%%
%%% Conflict of interest. ????????????
%%%%%%%%%%%%%%%%%%%%%%%%%%%%%%%%%%%%%%%%%%%%%%%%%%%%%%%
\InterestConflict{The authors declare that they have no conflict of interest.}

%%%%%%%%%%%%%%%%%%%%%%%%%%%%%%%%%%%%%%%%%%%%%%%%%%%%%%%
%%% Supplements. ????????, ????
%%%%%%%%%%%%%%%%%%%%%%%%%%%%%%%%%%%%%%%%%%%%%%%%%%%%%%%
%\Supplements{}

%%%%%%%%%%%%%%%%%%%%%%%%%%%%%%%%%%%%%%%%%%%%%%%%%%%%%%%
%%% Reference section. 
%%% citation in the content using "some words~\cite{1,2}".
%%% ~ is needed to make the reference number is on the same line with the word before it.
%%%%%%%%%%%%%%%%%%%%%%%%%%%%%%%%%%%%%%%%%%%%%%%%%%%%%%%

\end{multicols}
\end{document}